\documentclass[aps,prd,twocolumn,groupedaddress,amsmath,amssymb,superscriptaddress,showpacs]{revtex4}
\usepackage{mathrsfs, epsfig, graphicx, url, dcolumn, bm}

\begin{document}

\title{Cosmological Constraints from 21cm Surveys After Reionization}

\author{Eli Visbal}
\email[]{evisbal@fas.harvard.edu}
\affiliation{Jefferson Laboratory of Physics, Harvard University, Cambridge, MA 02138}
\affiliation{Harvard-Smithsonian CfA, 60 Garden Street, Cambridge, MA 02138}

\author{Abraham Loeb}
\affiliation{Harvard-Smithsonian CfA, 60 Garden Street, Cambridge, MA 02138}

\author{Stuart Wyithe}
\affiliation{School of Physics, University of Melbourne, Parkville, Victoria, Australia}


\date{\today}
\begin{abstract}
21cm emission from residual neutral hydrogen after the epoch of
reionization can be used to trace the cosmological power spectrum of
density fluctuations.  Using a Fisher matrix formulation, we provide a
detailed forecast of the constraints on cosmological parameters that
are achievable with this probe.  We consider two designs: a scaled-up
version of the MWA observatory as well as a Fast Fourier Transform
Telescope.  We find that 21cm observations dedicated to
post-reionization redshifts may yield significantly better constraints
than next generation Cosmic Microwave Background (CMB) experiments.
We find the constraints on $\Omega_\Lambda$, $\Omega_{\rm m}h^2$, and
$\Omega_\nu h^2$ to be the strongest, each improved by at least an order
of magnitude over the Planck CMB satellite alone for both designs.  
Our results do not depend as strongly on uncertainties in the
astrophysics associated with the ionization of hydrogen as similar 
21cm surveys during the epoch of reionization.  However, we find that
modulation of the 21cm power spectrum from the
ionizing background could potentially degrade constraints on the 
spectral index of the primordial power spectrum  and its running
by more than an order of magnitude.  Our results also depend
strongly on the maximum wavenumber of the power spectrum which can be
used due to non-linearities.
\end{abstract}


\maketitle


\section{Introduction}
Recently, there has been much interest in the feasibility of mapping
the three-dimensional distribution of cosmic hydrogen through its
spin-flip transition at a resonant rest frame wavelength of
21cm \cite{Furlanetto:2006jb,2007RPPh...70..627B}.  Several first
generation experiments are being constructed to probe the epoch of
reionization (MWA \cite{MWAref}, LOFAR \cite{LOFARref},
PAPER \cite{PAPERref}, 21CMA \cite{21CMAref}) and more ambitious designs
are being planned (SKA \cite{SKAref}).

One driver for mapping hydrogen through 21cm emission is to measure
cosmological parameters from the underlying cosmic power spectrum.
During the epoch of reionization (EoR), the 21cm power spectrum is
shaped mainly by the structure of ionized regions.  Even without
precise knowledge of the ionization power spectrum it is possible to
isolate the cosmological power spectrum by exploiting anisotropies in
redshift space due to peculiar velocities
 \cite{Barkana:2004zy,2006ApJ...653..815M}.
  
Recent work \cite{2008PhRvD..78b3529M,2008arXiv0805.1920P} has shown
that the 21cm power spectrum accessible during the EoR has the
potential to put tight constraints on cosmological parameters;
however, these constraints depend on model-dependent uncertainties,
the most important of which involves the ionization power spectrum.

In this paper, we explore in detail the constraints achievable from
mapping the residual cosmic hydrogen after reionization \cite{2007arXiv0708.3392W,2007arXiv0709.2955W,Chang:2007xk,2008PhRvL.100p1301L,Pen:2008fw,2008arXiv0808.2323W}.
The hydrogen resides in pockets of dense galactic 
regions which are self-shielded from the UV background, also known as damped
Lyman-$\alpha$ absorbers (DLAs) in quasar spectra \cite{Wolfe:2005jd}.
In difference from traditional galaxy redshift surveys, 21cm surveys
do not need to resolve individual galaxies but rather are able to monitor the
smooth variation in their cumulative 21cm emission owing to their
clustering on large scales \cite{2008PhRvL.100p1301L}.

Measuring the 21cm power spectrum after reionization offers several
key advantages as compared to the EoR.  First, the near uniformity of
the UV radiation field guarantees that the 21cm power spectrum would
reliably trace the underlying matter power spectrum.  
However, the UV background
could introduce a scale dependent modulation to the 
21cm power spectrum as large as
one percent \cite{2008arXiv0808.2323W}.  This 
modulation could introduce degeneracies that would significantly 
weaken constraints on the spectral index 
of the primordial power spectrum and its running.
Another advantage of probing low redshifts is that the
brightness temperature of the galactic synchrotron emission scales as
$(1+z)^{2.6}$.  However, this advantage is offset by the fact that the
mass weighted neutral fraction of hydrogen is only a few percent at
redshifts $z\lesssim 6$ \cite{Prochaska:2005wy}.

In this paper we use the Fisher matrix formalism to quantify how
effectively futuristic surveys dedicated to post-reionization
redshifts ($z\lesssim6$) can constrain cosmology.  We consider both a
survey with ten times the number of cross dipoles of the MWA \cite{MWAref} ($5000\times
16=8\times 10^4$) but operating at higher frequencies, which we term
MWA5000, and a Fast Fourier Transform Telescope (FFTT) with $10^6$
dipoles over a square kilometer area \cite{2008arXiv0805.4414T}.  
We also show results for the combination of these surveys and a next
generation Cosmic Microwave Background (CMB) experiment (Planck).

This paper is structured as follows. In \S II we discuss the detectability of 
the 21cm signal after reionization and in \S III we discuss its 
power spectrum. In \S IV we consider the details
of our Fisher matrix calculation and in \S V we present its results.
Finally, we discuss and summarize our conclusions in \S VI and
\S VII.

\section{Detectability of the 21cm Signal After Reionization}
Reionization starts with ionized (HII) regions around galaxies which
grow and eventually overlap.  This overlap of HII regions
characterizes the end of the EoR.  It was once thought that the 21cm
signal would disappear after this transition because there is little
neutral hydrogen left.  Recent work has shown that this is not the
case \cite{Wyithe:2008th,2007arXiv0708.3392W,2008PhRvL.100p1301L,
2008arXiv0808.2323W,Pritchard:2008da}.  The {\it detectability} of the
signal may not decline substantially following the end of the EoR
because the Galactic synchrotron foreground is weaker at higher
frequencies (lower redshifts of the 21cm emission).

After reionization most of the remaining neutral hydrogen is expected
to reside in DLAs.  Observations have shown that out to $z \approx 4$
the cosmological density parameter of HI is $\Omega_{\rm HI} \approx
10^{-3}$ \cite{Prochaska:2005wy}.  This corresponds to a mass-averaged
neutral fraction of a few percent. It does not contribute
significantly to the Ly$\alpha$ forest, which is mostly shaped by the
much smaller volume-averaged neutral fraction.  A 21cm survey like the
ones we discuss are sensitive to all neutral hydrogen within the
survey volume because no galaxies are individually identified and thus
there is no minimum threshold for the detection of individual
galaxies.

Even though the majority of the neutral hydrogen resides in self
shielded clumps, self absorption is not expected to significantly
reduce the 21cm signal.  This is supported by 21cm absorption studies
of DLAs over the redshift interval $0<z \lesssim 3.4$, which exhibit
an optical depth to 21cm absorption of the radio flux from a
background quasar of less than a few percent
\cite{Kanekar:2002sv,Curran:2007nk}.  These observations are supported
by theoretical calculations of the 21cm optical depth of neutral gas
in high redshift minihalos \cite{Furlanetto:2002ng}.  We also note
that the spin temperature in DLAs is much higher than the CMB
temperature and thus the 21cm signal is independent of the kinetic
temperature of the gas \cite{Kanekar:2002sv}.

At $z \approx 4$ the 21cm brightness temperature contrast with the CMB
will be roughly $0.5$ mK.  On scales of 10 comoving Mpc, the
\emph{rms} amplitude of density fluctuations is roughly
$\sigma\approx0.2$, so we expect 21cm fluctuations at a level of
$\approx 0.1$mK.  This is only an order of magnitude or so less than
the largest fluctuations expected during the EoR
\cite{2007astro.ph..3070W}.  These fluctuations combined with the
lower brightness temperature of galactic synchrotron emission, which
scales as $(1+z)^{2.6}$, should provide a detectable signal.  Previous
work (e.g. Fig. 10 in Ref.\cite{Pritchard:2008da}) shows that the
signal to noise ratio of the 21cm signal after reionization may be
similar or even higher than that during the EoR.  However, it is
important to note that it will be necessary to build special
instruments which are optimized to these redshifts to obtain the type
of results discussed in this paper.

\section{21cm Power Spectrum}

Next, we describe the 21cm power spectrum after the epoch
of reionization.  A complete discussion of the relevant physics can be
found in Refs.\ \cite{Furlanetto:2006jb,2007arXiv0708.3392W,2008PhRvL.100p1301L,2008arXiv0808.2323W}.  
The difference between the average
brightness temperature of 21cm emission at redshift $z$ 
and the CMB temperature is described by
\begin{equation}
\bar{T}_{\rm b} \approx 26\bar{x}_{\rm H} \left (
\frac{\bar{T}_{\rm s}-T_{\rm CMB}}{\bar{T}_{\rm s}} \frac{\Omega_{\rm b}h^2}{0.022} \right)
\left(\frac{0.15}{\Omega_{\rm m}h^2} \frac{1+z}{10} \right)^{1/2} {\rm mK},
\end{equation}
where $\bar{x}_{\rm H}$ is the global neutral hydrogen fraction and $T_{\rm s}$ is
the HI spin temperature.
In Fourier space, the power spectrum of 21cm
brightness fluctuations is defined by 
$(2\pi)^3 \delta^3({\bf k-k'})  P_{21}({\bf k}) \equiv \langle \Delta T_{\rm b}(\mathbf{k})^* \Delta T_{\rm b}
(\mathbf{k'})
\rangle$, where $\mathbf{k}$ is the wave-vector of a given Fourier mode and $\Delta T_{\rm b}(\mathbf{k})$
is the brightness temperature fluctuation in Fourier space.  

Before giving its exact form, we stress that the most important
feature of $P_{21}({\bf k})$ after reionization is that it essentially
traces the cosmological matter power spectrum, $P(k)$.  Extracting the
cosmological power spectrum will make it possible to place tight
constraints on many cosmological parameters.  During the EoR the
situation is more complicated.  $P_{21}({\bf k})$ contains additional
terms which depend on the ionization power spectrum and the cross
power-spectrum between the ionization and matter distributions.
  
Following the derivation of Ref. \cite{2008arXiv0808.2323W} 
we find that after reionization
\begin{equation}
\label{PSeqn}
P_{21}(\mathbf{k}) = \tilde{T^2_{\rm b}}\bar{x}_{\rm HI}^2P(k)\left[B(\mathbf{k}) +
f\mu^2\right]^2. 
\end{equation}
Here  $\tilde{T}_{\rm b} = \bar{T}_{\rm b}/(\bar{x}_{\rm HI}(\bar{T}_{\rm s}-T_{\rm 
CMB})/T_{\rm S})$,
$B(\mathbf{k})$ is the average scale dependent, $\bar{x}_{\rm HI}$ mass-weighted bias for the DLA host galaxies,
and $\mu=\cos\theta$ where $\theta$ is the angle between the line of sight
and the wave-vector $\mathbf{k}$
 \cite{Wyithe:2008th,2008arXiv0808.2323W}.  
The factor $[B(\mathbf{k}) + f\mu^2]^2$ 
arises due to redshift space anisotropies from line of sight peculiar
velocities \cite{1987MNRAS.227....1K,Bharadwaj:2004nr,Barkana:2004zy}.
Here the growth index is
defined as $f=d\ln D /d \ln a$, where $D(z)$ is the growth factor of
density perturbations and $a$ is the cosmic scale factor.  The bias
factor arises because the residual neutral hydrogen is located in host
galaxies which are biased with respect to the underlying dark matter
distribution.  This bias and the neutral hydrogen fraction are the
only quantities in the above equation which do not depend solely on
fundamental physics.

While the near uniformity of the ionizing UV background
causes $P_{21}({\bf k})$ to trace the cosmological power spectrum, 
fluctuations introduce a modulation that could modify the
power spectrum by less than one percent  \cite{2008arXiv0808.2323W}.
This modulation of the power spectrum
adds a scale dependent correction factor to the host
galaxy bias.  
The average bias $B(\mathbf{k})$ is equal to the 
mean halo bias $\bar{b}$, multiplied by a factor
$(1-K(k))$.  In appendix A, we derive an expression
for $K(k)$ for the case where the mean-free-path of ionizing photons
is independent of position.  If the effect of the ionizing
background is near the maximum which could be expected it 
may be necessary to model it in order to estimate cosmological parameters
reliably.  In \S V, we show how cosmological constraints would
likely be affected by degeneracies with nuisance parameters
introduced by a scale dependent bias model.



The bias, $ b$, for halos of a particular mass $M$ can be derived using the Press-Schechter formalism
modified to include non-spherical collapse \cite{2001MNRAS.323....1S}
\begin{multline}
\label{bias1}
b(M,z)=1 + \\
 \frac{1}{\delta_{\rm c}}\left[\nu'^{2}
  +b\nu'^{2(1-c)}-\frac{\nu'^{2c}/\sqrt{a}}{\nu'^{2c}+b(1-c)(1-c/2)}\right],
\end{multline}
where $\nu'^2=a\delta^2_{\rm c}/\sigma^2(M,z)$, $a=0.707$, $b=0.5$, and $c=0.6$.  
Here $\sigma^2(M,z)$ is the variance of the density field smoothed on
a mass scale M and $\delta_{\rm c}$ is the linear over-density 
threshold for collapse at redshift $z$.  

Assuming that the neutral gas to halo mass ratio is independent of mass, we can 
derive the mean bias at a particular redshift, $\bar{b}$, as the weighted average
\begin{equation}
\label{bias2}
\bar{b}=\frac{\int_{M_{\rm min}}^\infty \frac{dn}{dM}b(M,z)MdM}{\int_{M_{\rm min}}^\infty \frac{dn}{dM}MdM},
\end{equation}
where $dn/dM$ is the Sheth-Tormen mass function of dark matter halos \cite{2001MNRAS.323....1S}.
The minimum halo mass $M_{\rm min}$ defines the threshold 
for assembling heated gas out of the photo-ionized intergalactic medium,
corresponding to a minimum virial temperature $T_{\rm vir} \approx 10^5 K$
 \cite{Wyithe:2006st,1992MNRAS.256P..43E,1994ApJ...427...25S,1996ApJ...465..608T,1997MNRAS.292...27H,2008MNRAS.390.1071M}.
The appropriate value of $M_{\rm min}$ will depend on 
the mass of galaxies which host DLAs, however we find our
results insensitive to the exact choice of $M_{\rm min}$ (see Fig. \ref{shot}).
   
A constant virial temperature leads to the redshift dependence
$M_{\rm min}=5.3\times10^9\left((1+z)/{4.5}\right)^{-1.5} M_\odot$.
Based on observations of DLAs \cite{Prochaska:2005wy}, we assume a fiducial value of
neutral hydrogen density to be $x_{\rm HI1}=0.02$.

Previous work has suggested that at high redshifts the galaxy power spectrum 
may be affected by reionization \cite{Babich:2005jj,2007arXiv0706.3744W}.  
In \cite{Babich:2005jj} it is shown that 
fluctuations in the redshift of reionization in different locations
leads to variation in the minimum mass of galaxies which can be assembled.  
This creates a scale dependent modulation of the power spectrum.
Recent observations find that at $z\approx3$ galaxies with virial
velocities $< 50-70 {\rm km/s}$ contribute 
little to DLAs and that most DLAs are
hosted within halos having viral velocities between $V=100-150$ km/s \cite{2009MNRAS.tmp..820B}.
Since DLAs and thus most of the remaining neutral hydrogen after reionization 
is not found in the smallest galaxies we do not expect changes in the
 minimum mass of galaxies which are assembled to have an 
important effect on the 21cm power spectrum.

In \cite{2007arXiv0706.3744W}, the effects of 
reionization on the galaxy power spectrum 
for more massive galaxies are explored.  Reionization is found to change 
the power spectrum, but this was due to reionization modifying the
average stellar age of galaxies and hence the mass to light ratio. 
We do not expect this to significantly effect the neutral hydrogen 
hosted in galaxies.  Additionally the modulation of the power spectrum
was found to be approximately scale independent, which would not be
degenerate with cosmological parameters discussed in this work.

\subsection{Non-Linear Effects on Baryon Acoustic Oscillations}
It has been recently shown that non-linear effects may change the
baryon acoustic oscillations (BAO) signature at larger scales than those
previously considered \cite{2007ApJ...665...14S}.  The erasure of BAO
information due to non-linear effects can be written as
\begin{multline}
\label{NL}
P_{\rm b,nl}(k,\mu) = P_{\rm b,lin}(k) \times  \\
\exp \left [ -\frac{k^2}{2}
\left((1-\mu^2)\Sigma^2_\perp + \mu^2\Sigma^2_\| \right) \right ],
\end{multline}
where $P_{\rm b}$ is the part of the power spectrum which contains the
wiggles from the BAO, $\Sigma_\perp=\Sigma_0{D}(z)$,
$\Sigma_\|=\Sigma_0(1+f){D}(z)$,  $\Sigma_0=11.6h^{-1}{\rm Mpc}$, and 
${D}(z)$ is the growth function normalized to $(1+z)^{-1}$
at high redshifts.  In our Fisher matrix calculations 
$(P_{\rm b,nl}(k)+P_{\rm smooth}(k))$  
is used in place of $P(k)$ in Eq. \ref{PSeqn}.
$P_{\rm b,lin}(k)=P(k)-P_{\rm smooth}(k)$, where
$P_{\rm smooth}(k)$ is a low order polynomial fit to $P(k)$ in log-log
space.  This fit removes the wiggles from the BAO, such that $P_{\rm b}$
is just the part of the power spectrum containing the BAO wiggles.
Our choice of $\Sigma_0$ corresponds to the value determined by
simulations and scaled linearly with $\sigma_8$ to fit our fiducial
cosmology.  The scales implied by $\Sigma_\perp$ and $\Sigma_\|$  
are held fixed in observable units such that the exponential suppression 
can be taken outside of the derivatives in the Fisher matrix formalism 
described below.  This ensures that the BAO erasure does not in any way 
improve the forecasted errors on cosmological parameters.

We emphasize that the exponential suppression of the power spectrum
in Eq. \ref{NL} is only applied to the the BAO wiggles after the smooth 
linear power has been subtracted off.  The power spectrum used in the
derivatives which enter the Fisher matrix described below
 is the sum of the wiggles suppressed
by Eq. \ref{NL} and the unsuppressed smooth linear power spectrum.  For 
the error on $P_{\rm 21}$ which enters the Fisher matrix we do 
not suppress either the wiggles or the 
smooth part of the power spectrum.

\section{Fisher Matrix Formulation}
Given a set of parameters $\lambda_i$, the Fisher matrix formalism
provides an estimation of the error for each of the parameters
associated with some data set \cite{1997ApJ...480...22T}.
The $1\sigma $ errors on parameters can be estimated as
\begin{equation}
 \Delta\lambda_i = \sqrt{F^{-1}_{ii}},
\end{equation}
where
\begin{equation}
\label{FISHEReqn}
F_{ij}=\sum_{pixels}\frac{1}{\delta P_{12}^2} \left( \frac{\partial
P_{21}}{\partial \lambda_i} \right) \left( \frac{\partial
  P_{21}}{\partial\lambda_j} \right),
\end{equation}
 \cite{2006ApJ...653..815M,2008PhRvD..78b3529M}, $P_{21}$ is the
total 21cm power spectrum, and $\delta P_{21}$ is the uncertainty on a
measurement of the power spectrum.  The relevant derivatives have been
calculated using the transfer functions from CAMB \cite{CAMBref}.

We work in $\mathbf{u}$-space rather than $\mathbf{k}$-space as described in
Refs.\ \cite{2008PhRvD..78b3529M,2006ApJ...653..815M} to simplify calculating Alcock-Paczynski
effects.  A radio interferometer 
directly measures visibilities, 
\begin{equation}
V(u,v,\nu) = \int d{\bf{\hat{n}}} \Delta T_{\rm b}({\bf{\hat{n}}},\nu) A_\nu({\bf{\hat{n}}})e^{2\pi i (u,v) \cdot {\bf{\hat{n}}}},
\end{equation}
where $V$ is the visibility for a pair of antennae and $A_{\nu}$ is the 
contribution to the primary beam in the $\mathbf{\hat{n}}$ direction.  
Here we have used the flat sky approximation.  This is appropriate even in the case of 
the FFTT, which images the entire hemisphere, because essentially 
all of the cosmological information in our surveys is found on small angular scales.    
The vector $\mathbf{u_\perp}=(u,v)$ corresponds to the number of 
wavelengths between the pair of antennae.  Performing the Fourier 
transform $I(\mathbf{u})=\int d\nu V(u,v,\nu) \exp(2\pi i \nu \eta)$, we obtain
a signal in terms of $\mathbf{u}=u\mathbf{\hat{i}}+v\mathbf{\hat{j}}+\eta\mathbf{\hat{k}}$, where $\eta$ 
has units of time and $\mathbf{\hat{k}}$ is the unit vector along the line of sight.
Note that there is a one to one correspondence between ${\bf k}$ and ${\bf u}$ given
by $2\pi{\bf u}_\perp/d_A={\bf k}_\perp$ perpendicular to the line of sight and 
$2\pi{\bf u}_\|/\tilde{y}={\bf k}_\|$ along the line of sight, 
where $d_A$ is the angular diameter distance to the observation and $\tilde{y}$ is the ratio
of comoving distance to frequency interval.
With the Fourier conventions above 
we have $P_{21}({\bf k})=P_{21}({\bf u})d_A^2 \Tilde{y}$. 
Working in $\mathbf{u}$-space simplifies our calculation because
$P_{21}({\bf u})$ is measurable without cosmological assumptions and
thus Eq.\ (\ref{FISHEReqn}) can be applied directly.  If we were to work
in $\mathbf{k}$-space, the Alcock-Paczynski effect would distort
$P_{21}({\bf k})$ from Eq.\ (\ref{PSeqn}) when we take derivatives with
respect to cosmological parameters.
  
Because we are considering discrete sources, the total 21cm power
spectrum will be the sum of Eq.\ (\ref{PSeqn}) 
and a shot noise term $P_{\rm shot}=P_{21}(k)/(\bar{b}^2P(k)n_{\rm DLA})$,
where $n_{\rm DLA}$ is the effective number density of 
the galaxies which host DLAs, 
$P_{21}(k)$ is $P_{21}(\mathbf{k})$ along the line of sight, and $\bar{b}^2P(k)$
is the power spectrum of DLA hosting galaxies along the line of sight.

Any constant $|{\mathbf{k}}|$ and $\theta$
define an annulus of constant $P_{21}(\mathbf{u})$.
The Fisher matrix
is calculated from Eq.\ (\ref{FISHEReqn}) by summing the
contribution from annuli which fill all of ${\bf k}$-space accessible to an observation.  
We divide a 21cm survey into redshift bins small 
enough for the redshift evolution of
$P(k)$ across a bin to be negligible.  A separate Fisher matrix is
calculated for each redshift bin and then summed into a Fisher matrix
which reflects the information about the entire survey.

To calculate the error in Eq.\ (\ref{FISHEReqn}) 
we follow the work of Refs. \cite{2006ApJ...653..815M,2005ApJ...619..678M}.
The error on $P_{21}$ for a particular $|{\mathbf{k}}|$  and $\theta$ is 
\begin{equation}
\delta P_{21} = \frac{P_{21} + P_{\rm shot} + P_{N}}{\sqrt{N_{\rm c} \times N_{\rm fields} \times B_{\rm tot}/B}},
\end{equation}
where
$P_{N}$ is the noise power spectrum of an interferometer, $B$ is the bandwidth
over which foregrounds can be removed, $B_{\rm tot}$ is the total
accessible bandwidth, $N_{\rm fields}$ is the number of fields being imaged, and
$N_{\rm c}$ is the number of independent cells in ${\bf k}$-space. 
The number of cells in an annulus of constant $P_{21}(\mathbf{u})$ is well 
approximated by

\begin{equation}
N_{\rm c} = 2\pi k^2 \sin \theta \Delta k \Delta \theta \frac{V}{(2\pi)^3},
\end{equation}
where $V$ is the comoving volume observed by the experiment, $\Delta
\theta$ and $\Delta k$ are the angular and ${\bf k}$-space widths of the
annulus.  We choose $\Delta k$ and $\Delta \theta$ such that $P_{21}(\mathbf{u})$
is essentially constant. 
The noise of an interferometer is given by
\begin{equation}
P_{\rm N}(u_\perp) = \left(\frac{\lambda^2T_{\rm sys}}{A_{\rm e}}
\right)^2\frac{1}{t_0n(u_\perp)},
\end{equation}
where $\lambda=21{\rm cm}\times (1+z)$ is the observed wavelength,
$T_{\rm sys}$ is the system temperature of the interferometer, $A_{\rm e}$ is
the effective area, $t_0$ is the total observing time, and $n(u_\perp)$
is the number density of baselines.  This error is identical to that
applicable in studies of the EoR with the exception of the shot noise
\cite{2008PhRvD..78b3529M,2006ApJ...653..815M}.  

The sensitivity of 21cm measurements will be impacted by how
effectively foregrounds such as the galactic synchrotron emission can
be removed.
Refs. \cite{Wang:2005zj,2006ApJ...653..815M,2008MNRAS.389.1319J}
suggest that after fitting out a low order polynomial, residual
foregrounds in the power spectrum will be negligible if the frequency
band over which foregrounds are removed, $B$, is substantially smaller
than the total band-pass available.  This is true for wave-vectors
greater than $k_{\rm min} = 2\pi/(\tilde{y}B)$.  Recent work
estimating errors on cosmological parameters with 21cm surveys during
the EoR has used this assumption in modeling foreground removal
\cite{2008PhRvD..78b3529M,2008arXiv0805.1920P,2006ApJ...653..815M}.
It is encouraging that we find parameter constraints to depend
relatively weakly on the exact value of the bandwidth over which
foregrounds are removed (see Fig. \ref{BW}).  We note that while we
expect these assumptions pertaining to foreground removal to be
robust, the true ability to remove foregrounds may not be known until
first generation 21cm instruments are operational.

There are several factors which limit the scales accessible to 21cm
surveys for parameter determination.  As discussed above, we assume
that foregrounds can not be effectively removed from pixels with $k <
k_{\rm min}$.  There is also a minimum accessible value of
$k\sin\theta$ imposed by the minimum baseline, $b_{\rm min} \approx
\sqrt{A_{\rm phys}}$.  We exclude information for $k$ values above the
linear scale.  We compare the linear power spectrum for our fiducial
cosmology to a non-linear power spectrum produced by HALOFIT in CAMB
\cite{CAMBref}, and define our non-linear cutoff as the $k$ value
where there is a $10 \%$ discrepancy.  

The precision of the measurements
discussed in this paper will require a theoretical prediction of the matter
power spectrum which is accurate to better than one percent.  The $k$
cutoff is meant to represent the scale to which future N-body simulations,
such as those discussed in \cite{2008arXiv0812.1052H,2009arXiv0902.0429H}, 
will be able to calculate the 
power spectrum to the required accuracy.     
Since the exact value of this 
scale is unknown we present the dependence of varying it on cosmological
parameter constraints in Fig. \ref{kmax}. 

\begin{figure}
\includegraphics{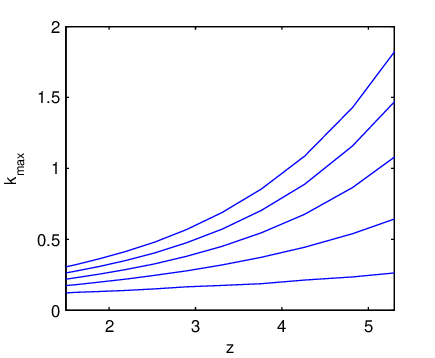}
\caption{\label{cut} Maximal wavenumber, $k_{\rm max}$, above which the
non-linear power-spectrum deviates from the linear power-spectrum by
more than $25\%,20\%,15\%,10\%$ and $5\%$ (top to bottom).
Larger discrepancies produce higher values of $k_{\rm max}$.}
\end{figure}

The value of this cutoff is
fairly conservative when compared to the criterion of Ref.\
\cite{Seo:2003pu} which chooses $k_{\rm nl}=\pi/2R$ for a value of $R$
giving an {\it rms} density fluctuation amplitude $\sigma(R) = 0.5$.
The latter criterion has been used for 21cm parameter constraints at
higher redshifts than we explore \cite{2008arXiv0805.1920P}.
The $k_{\rm max}$ values corresponding to different levels of discrepancy 
between the linear and non-linear power spectra are shown in Fig.\ \ref{cut}. 

We also include constraints when combined with 
future data from the Planck satellite \cite{Planckref}.  
The method used to calculate the corresponding Fisher matrix 
was adopted from Refs.\ \cite{Jungman:1995bz,Zaldarriaga:1996xe}.

\subsection{Survey Characteristics}

We apply the Fisher matrix formulation to future surveys dedicated to
low redshifts similar in design to MWA5000 and a Fast Fourier
Transform Telescope (FFTT).  We show results with surveys designed for
a central redshift of 3.5 and 1.5 that each span a factor of 3 in
$(1+z)$.

We model the MWA5000 observatory as an interferometer with
5000 tiles each containing 16 dipole antennae. 
For the survey centered at $z=3.5(1.5)$ we assume a constant core of
antennae out to 40(22)m and then a $r^{-2}$ distribution out to
580(278)m.  This layout gives roughly the same baseline density
distribution as MWA does at higher redshifts
\cite{2006ApJ...653..815M}.  

The angular resolution of the telescope
is set by the baseline density distribution.  The longest baseline
corresponds to the highest value of $k_\perp$ that can be probed,
$2\pi{\bf u}_{\perp,max}/d_A={\bf k}_{\perp,max}$.  
The above distribution of antennae contains baselines
which are long enough to resolve $k_{max}$ at all redshifts.  
We assume a frequency resolution of 0.01 MHz for all 
surveys considered.  These arrays will
have better resolution than this, but this is sufficient 
to resolve the scales set by $k_{max}$.  

For the survey centered around $z=3.5$ the effective area of each tile
$A_{\rm e} \approx N_{\rm dip}\lambda^2/4$ and is limited by the physical area
of the tile, $A_{\rm phys}$ \cite{Bowman:2005cr}.  The values of
$A_{\rm phys}$ are chosen such that $A_{\rm phys} \approx A_{\rm e}$ at the
central redshift.  For the low-redshift case, we assume that dishes
are used instead of antennae, resulting in $A_{\rm e} \approx A_{\rm phys}$.

We model an FFTT observatory \cite{2008arXiv0805.4414T} as an
interferometer with $10^6$ evenly spaced dipoles over a square
kilometer that can all be correlated.  The noise in each dipole is
calculated separately with $A_{\rm e} \approx \lambda^2/4$ limited by
$A_{\rm phys} = S^2$, where $S$ is the spacing of the dipoles on a square
grid.

We have assumed that foregrounds can be removed on scales of up to
$30[4.5/(1+z)]$MHz.  This corresponds to the value that \cite{2008PhRvL.100p1301L} cite as 
desirable to measure the neutrino signal.  The exact choice of 
this bandwidth is relatively unimportant, as our results depend weakly on 
this value (see Fig.\ \ref{BW}).  
We assume that 0.65 of the sky in one hemisphere is imaged for
2000 hours.  This corresponds to roughly 16 fields of view.  
The partial coverage is due to foregrounds which can not
be removed in the vicinity of the galactic plane.  We also show
results for a more conservative case using MWA5000 over three fields of view rather than
16.

For all surveys we assume a redshift range corresponding to a factor
of 3 in frequency and break each observation into redshift bins of
width $\Delta z = 0.1 (1+z)$.  The factor of 3 corresponds to the
largest frequency bandwidth over which a low frequency dipole antenna
has suitable sensitivity.  In our lower redshift case we have assumed
that dishes will be used.  These could potentially have a larger
bandwidth than the dipole antennae, but we conservatively use the same
factor of 3 in frequency. We also assume that $T_{\rm sys}=T_{\rm sky} +
T_{\rm inst}$ where the sky temperature $T_{\rm sky}=260 \left[ ({1+z})/{9.5}
\right]^{2.55}$ \cite{Rogers:2008vh}, and that the instrumental
temperature is 30K in an optimistic case and 100K in a pessimistic
case, corresponding to what is reasonable with current technology and
what might be expected by the time the hypothetical experiments we
discuss will be built.

\subsection{Cosmological Parameters}

Throughout this work we assume that the true background cosmology is
that of a flat $\Lambda$CDM Universe with density parameters
$\Omega_\Lambda=0.7$ in dark energy, $\Omega_{\rm m}h^2=0.147$ in matter,
$\Omega_{\rm b}h^2=0.023$ in baryons, and $\Omega_\nu h^2=0.00054$ in neutrinos, 
and with values of $A_{\rm s}^2=25.0 \times 10^{-10}$, $n_{\rm s}=0.95$, and
$\alpha=0.0$ for the spectral index, amplitude, and running of the
primordial power spectrum.  The chosen value of $A_{\rm s}$ corresponds to
an {\it rms} fluctuation amplitude on a $8h^{-1}$ Mpc scale of
$\sigma_8=0.84$.  We have assumed a neutrino hierarchy with one
dominant species having a neutrino mass of $0.05$eV.  We assume a
fiducial value of $\tau=0.1$ for the optical depth to electron
scattering during reionization, and a helium mass fraction from big
bang nucleosynthesis of $Y_{\rm He}=0.24$.  In a lower redshift example we
also leave the dark energy equation of state as a free parameter with
a fiducial value of $w=-1$.



%

\section{Results}

We present the results of our Fisher matrix calculations in Tables
\ref{min35}-\ref{atanpoly}. 
We consider several different sets of assumptions regarding the
effects of the ionizing background, which as explained above,
may introduce a weakly scale dependent bias.  In principle this could
produce degeneracies which would weaken constraints on cosmological
parameters.  First we present the case where the scale dependence of
the bias is insignificant compared with the precision of the power
spectrum measurements.  Then we assume that the effect is large enough
to require using a model with additional parameters to accurately fit the data.
We also discuss a very pessimistic scenario where the form
of the bias is completely unknown.

We calculate constraints for both optimistic and pessimistic
antenna temperature, which are 30K and 100K respectively.  
We also show results for a cosmic variance
limited version of each survey where the detector noise is assumed to
make no contribution.  This case represents the best conditions
imaginable for constraining cosmological parameters given the finite
volume of space that is observed.  As it turns out, an FFTT is
nearly cosmic variance limited at the redshifts of interest here.
In this case, we only show results for the optimistic antenna
temperature case.

We consider a flat $\Lambda$CDM cosmology except
for one case centered around $z=1.5$ where we leave $w$ as a free
parameter.

\subsection{Constant Bias}
We begin with the case where the ionizing background does not 
introduce any significant modulation to the power spectrum.  Thus, the
bias, $B({\bf k})=\bar{b}$ can be taken as a 
constant in each redshift bin.  Note that the value of the bias
is different in each redshift bin, by constant we are referring to
its dependence on scale.  This
is the most optimistic model one could hope for.
Here we assume that the shot noise, neutral fraction and
the bias in each redshift bin are unknown.  Each is marginalized over as a free
parameter in the Fisher matrix.  We use Eqs.\ (\ref{bias1}) and
(\ref{bias2}) to determine the fiducial values of the bias.
The shot noise is determined from the Press-Schechter mass function
of dark matter halos
modified to include non-spherical collapse.
The fiducial value of the neutral fraction 
is set to $x_{\rm HI}=0.02$.  The results are presented in Tables
\ref{min35}-\ref{minw}.

\subsection{Scale Dependent Bias}
We consider the case where the fluctuations of the 21cm power spectrum
due to the ionizing background are large enough to require fitting
additional nuisance parameters.  We do not claim to derive the exact 
form of the modulation, instead we assume an idealized form and fit,
 and then see
how cosmological constraints are degraded as more flexibility is introduced.    
As discussed above
the scale dependent bias, $B({\bf k})$, is the product of the average
galaxy bias, $\bar{b}$ and $(1-K({\bf k}))$.  In \cite{2008arXiv0808.2323W} , $K({\bf k})$
is calculated for the case where all photons travel the same distance.
Here we consider the more realistic case of a constant
mean-free-path.  In this case, as calculated in 
appendix A, the modulation takes the form 
$K({\bf k})=K_0 \arctan(k\lambda_{\rm mfp})/\lambda_{\rm mfp}k$.  
We use the following approximation 
(given here in proper distance, but appearing in comoving distance
in $K({\bf k})$) for the 
fiducial values of the mean-free-path:
$\lambda_{\rm mfp} = 85\left((1+z)/{4} \right )^{-4} {\rm Mpc}$ 
 \cite{FaucherGiguere:2008rc}.
As before, in our Fisher
matrix calculations we marginalize over shot noise, neutral fraction and 
average bias in each redshift bin.  Additionally, we marginalize
over $K_0$ and $\lambda_{\rm mfp}$ in each redshift bin.  
Our results are shown in Tables \ref{atan35}-\ref{atan15}.

To see how constraints are affected as we increase the flexibility of the fit,
we consider a case with small corrections to the form used above.  Here
we assume that the modulation is of the form
\begin{equation}
 K({\bf k})=K_0 \arctan(k\lambda_{\rm mfp})/\lambda_{\rm mfp}k + c_1k + c_2k^2 + c_3k^3 ...
\end{equation}
where the $c_n$'s are set to fiducial values of zero and marginalized 
over in each redshift bin.  Aside from marginalizing over these new parameters,
the Fisher matrix calculation is done the same as described above.  

Table \ref{atanpoly} shows the results for FFTT for marginalizing over all of the $c_n$'s
up to degree n.  This illustrates how parameter constraints will be degraded
as there is more flexibility in the fit for $K({\bf k})$.  We find that most
cosmological parameters are not degenerate with scale dependent changes
to the bias.  However, the spectral index of the primordial power spectrum 
and its running are degraded significantly.  This is because a change in these
parameters smoothly modifies the 21cm power spectrum as a function of k, which 
mimics changes due to scale dependent bias parameters.  A change in other parameters
alters features in the power spectrum that can be distinguished from the scale
dependent bias.  For example, one might expect $\Omega_\nu h^2$ to be strongly degenerate as well,
since neutrinos change the power spectrum in a smooth way.  However, with our choice of
cosmological parameters, we hold $\Omega_m h^2$ constant when taking derivatives of 
 $\Omega_\nu h^2$.  This means that changing the energy density of neutrinos also 
changes the energy density of cold dark matter.  The associated change in the cold dark matter
component alters features in the power spectrum which break the
degeneracy with the scale dependent bias.

Note that while the above fit gives a range of 
how much fitting the scale dependent bias could weaken cosmological constraints,
future work will have to determine the exact form of the fit which should 
be used.  This will require detailed numerical simulations.

We have also considered the extremely pessimistic
case where the form of the scale dependent 
bias is completely unknown.  It has been found that
it may be possible to isolate the part of the power 
spectrum containing only cosmological information by
exploiting the angular dependence of the power spectrum.
We use the PESS ionization power spectrum model 
described in \cite{2008PhRvD..78b3529M} to calculate 
Fisher matrices.  This isolates the term in the 21cm power
spectrum which does not depend on the bias.
We find that even FFTT cannot constrain
cosmological effectively and does not improve constraints
from Planck significantly.  This is probably overly 
pessimistic since the changes to the bias from the ionizing
background are expected to be small, but it shows
the importance of understanding the form of the bias.

 \begin{figure}
 \includegraphics{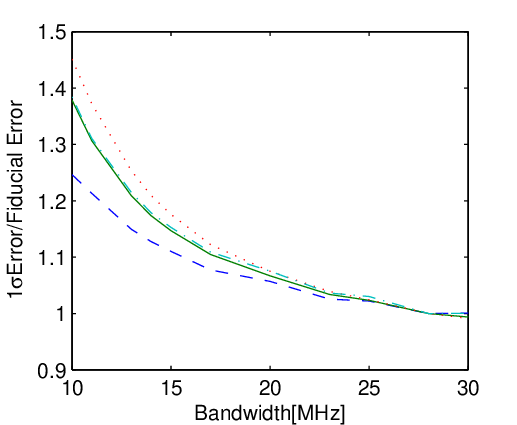}
 \caption{\label{BW} Sensitivity of the relative $1\sigma$ error bar
to the bandwidth over which foregrounds can be removed for
FFTT. Errors have been normalized such that they equal unity for the
assumptions in Table \ref{min35}.    The dashed line gives the error for $\Omega_\Lambda$, the
dotted line for $\Omega_\nu h^2$, the solid line for $n_{\rm s}$
 and the dot-dash line for $\alpha$.}
 \end{figure}

 \begin{figure}
 \includegraphics{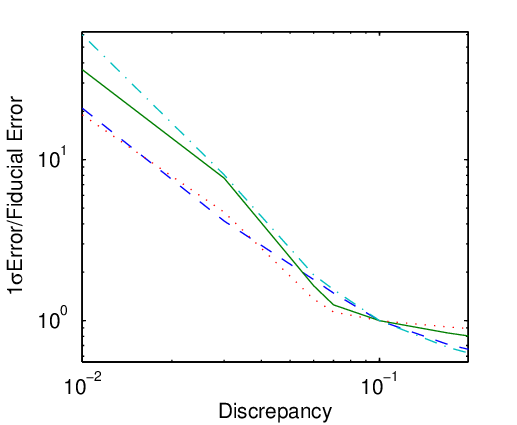}
 \caption{\label{kmax} Sensitivity of the relative $1\sigma$ error bar
to the maximal wavenumber $k_{\rm max}$ for FFTT.
Errors have been
normalized such that they equal one for the assumptions in Table
\ref{min35}.  The dashed line gives the error for $\Omega_\Lambda$, the
dotted line for $\Omega_\nu h^2$, the solid line for $n_{\rm s}$
 and the dot-dash line for $\alpha$.}

 \end{figure}

 \begin{figure}
 \includegraphics{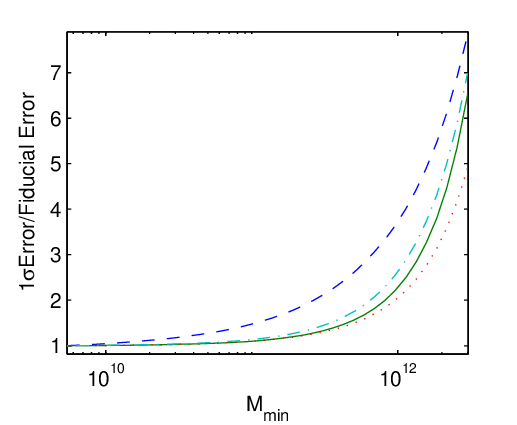}
 \caption{\label{shot} Dependence of constraints 
on $M_{min}$ (in units of solar mass) 
for FFTT.  Errors have been normalized such that they
equal one for the assumptions in table \ref{min35}.  The dashed line gives the error for $\Omega_\Lambda$, the
dotted line for $\Omega_\nu h^2$, the solid line for $n_{\rm s}$
 and the dot-dash line for $\alpha$.}  
 \end{figure}

\subsection{Other Factors}
There are several other factors that affect the constraints attainable
with post-reionization 21cm surveys.  We investigate how these factors
change our constraints in the case of a constant bias.  In Fig.\ \ref{BW}, we
have plotted the dependence of the $1\sigma$ uncertainty on the
bandwidth over which foregrounds are removed.  Larger values for the
bandwidth allow measurement of smaller $k$ values which are otherwise
not accessible.  Overall, the sensitivity to the bandwidth variation
is relatively modest.

In Fig.\ \ref{kmax}, we show how the constraints vary for different
choices of $k_{\rm max}$, the maximum wavenumber being used.  We show how
the parameter uncertainties change if we impose the cutoff at
different levels of discrepancy between the non-linear and linear
power spectra.  As $k_{\rm max}$ is lowered errors on parameters are
substantially increased.  For a discrepancy of 0.1 and 0.05, $k_{\rm max}$
ranges from $0.68-0.16 {\rm Mpc}^{-1}$ 
and $0.28-0.12 {\rm Mpc}^{-1}$ respectively,
see Fig.\ \ref{cut}.  For a discrepancy of 0.01 
$k_{\rm max}\approx 0.06 {\rm Mpc}^{-1}$ for each redshift bin. 

In Fig.\ \ref{shot}, we show how the constraints depend on $M_{\rm min}$,
the minimum halo mass above which significant amounts of neutral hydrogen
may assemble.  A high value of $M_{\rm min}$ increases the shot noise and
degrades the constraints.  We find that the shot noise only affects
the results significantly above $\sim 10^{12}M_\odot$, which is
unrealistic since hydrogen is found in lower mass galaxies 
in the local Universe \cite{Davies:2001ana}.


%



\begin{table*}
\tiny{
\caption{\label{min35} Constant Bias: 1$\sigma$ Errors on cosmological 
parameters with surveys centered at $z=3.5$.  OPT refers to an
antenna temperature of 30K, PESS 100K, and CVL to the 
cosmic variance limited case.  OPT and PESS correspond to what may
be possible by the time these experiments 
will be built and what is achievable today.}
\begin{ruledtabular}
\footnotetext[1]{When not otherwise indicated the ``OPT'' value of $T_{\rm sys}$ has been used.}
\begin{tabular}{lllllllllllll}
  & &     $\Omega_\Lambda$ & $\Omega_{\rm m}h^2$ & $\Omega_{\rm b}h^2$  & $n_{\rm s}$ & $A_{\rm s}^2 \times 10^{10}$ & $\alpha$ & $\Omega_\nu h^2$   & $\tau$ & $Y_{\rm He}$ \\ \hline
Fiducial Values     &     & 0.7  & 0.147  & 0.023  & 0.95   & 25.0     & 0.0     & 0.00054    & 0.10 & 0.24 \\ \hline
MWA5k-Hemisphere    &  PESS    &   0.0041 & 0.0098 & 0.0024 & 0.03 & -- & 0.012 & 0.0018 & -- & -- \\
                    &  OPT     &   0.0024 & 0.0041 & 0.001 & 0.014 & -- & 0.0068 & 0.00095 & -- & -- \\
                    &  CVL     &   0.00059 & 0.00025 & 0.000074 & 0.003 & -- & 0.0013 & 0.0003 & -- & -- \\ \hline
MWA5k-3 Fields      &  PESS    &   0.0095 & 0.023 & 0.0056 & 0.07 & -- & 0.029 & 0.0042 & -- & -- \\ 
                    &  OPT      &  0.0055 & 0.0096 & 0.0023 & 0.034 & -- & 0.016 & 0.0022 & -- & -- \\    
                    &  CVL      &  0.0014 & 0.00059 & 0.00017 & 0.007 & -- & 0.003 & 0.00071 & -- & -- \\      \hline
FFTT                &           &  0.00076 & 0.00032 & 0.000087 & 0.0034 & -- & 0.0016 & 0.00032 & -- & -- \\    \hline
Planck              &           &  0.038 & 0.0041  & 0.00024  & 0.0094  & 0.25   & 0.007  & 0.0039  & 0.0046   & 0.014  \\
$+$MWA5k-Hemisphere &           &  0.0023 & 0.00044 & 0.00013 & 0.0037 & 0.21 & 0.0028 & 0.00041 & 0.004 & 0.0057 \\
$+$MWA5k-3 Fields   &           &  0.0053 & 0.00074 & 0.00017 & 0.0056 & 0.22 & 0.004 & 0.00077 & 0.0041 & 0.0082 \\  
$+$FFTT             &           &  0.00076 & 0.00022 & 0.000076 & 0.0027 & 0.2 & 0.0013 & 0.00026 & 0.004 & 0.0043 \\
\end{tabular}          
\end{ruledtabular}  
}                            
\end{table*}

\begin{table*}
\tiny{
\caption{\label{min15} Constant Bias: 1$\sigma$ Errors on 
cosmological parameters with surveys centered at $z=1.5$.  
OPT refers to an
antenna temperature of 30K, PESS 100K, and 
CVL to the cosmic variance limited case.  
OPT and PESS correspond to what may
be possible by the time these experiments will 
be built and what is achievable today.}
\begin{ruledtabular}
\footnotetext[1]{When not otherwise indicated the ``OPT'' value of $T_{\rm sys}$ has been used.}
\begin{tabular}{lllllllllllll}
  & &     $\Omega_\Lambda$ & $\Omega_{\rm m}h^2$ & $\Omega_{\rm b}h^2$  & $n_{\rm s}$ & $A_{\rm s}^2 \times 10^{10}$ & $\alpha$ & $\Omega_\nu h^2$   & $\tau$ & $Y_{\rm He}$ \\ \hline
Fiducial Values     &           & 0.7      & 0.147    & 0.023    & 0.95   & 25.0     & 0.0     & 0.00054    & 0.10 & 0.24 \\ \hline
MWA5k-Hemisphere    &  PESS     &  0.0033 & 0.016 & 0.004 & 0.051 & -- & 0.019 & 0.0027 & -- & -- \\
                    &  OPT      &  0.0023 & 0.011 & 0.0027 & 0.036 & -- & 0.014 & 0.002 & -- & -- \\
                    &  CVL      &  0.0018 & 0.0085 & 0.0021 & 0.029 & -- & 0.011 & 0.0016 & -- & -- \\   \hline
MWA5k-3 Fields      &  PESS     &  0.0077 & 0.038 & 0.0093 & 0.12 & -- & 0.044 & 0.0063 & -- & -- \\
                    &  OPT      &  0.0054 & 0.026 & 0.0063 & 0.084 & -- & 0.032 & 0.0046 & -- & -- \\    
                    &  CVL      &  0.0042 & 0.02 & 0.0048 & 0.068 & -- & 0.026 & 0.0038 & -- & -- \\       \hline
Planck              &           &  0.038    & 0.0041   & 0.00024  & 0.0094  & 0.25   & 0.007  & 0.0039  & 0.0046   & 0.014  \\
$+$MWA5k-Hemisphere &           &  0.0021 & 0.00042 & 0.00014 & 0.0042 & 0.22 & 0.0033 & 0.00041 & 0.0041 & 0.0065 \\
$+$MWA5k-3 Fields   &           &  0.0047 & 0.00068 & 0.00018 & 0.0064 & 0.22 & 0.0048 & 0.00075 & 0.0041 & 0.0097  \\  
\end{tabular}                                                                                                                   
\end{ruledtabular}                                                      
}                    
\end{table*}

\begin{table*}
\tiny{
\caption{\label{plancktable} Ratio of 1$\sigma$ Errors for surveys combined with Planck to Errors from Planck alone.}
\begin{ruledtabular}
\footnotetext[1]{The ``OPT'' value of $T_{\rm sys}$ (30K) and the constant bias model have been used.}
\begin{tabular}{lllllllllllllll}
  & &     $\Omega_\Lambda$ & $\Omega_{\rm m}h^2$ & $\Omega_{\rm b}h^2$  & $n_{\rm s}$ & $A_{\rm s}^2$ & $\alpha$ & $\Omega_\nu h^2$   &$\tau$ & $Y_{\rm He}$\\ \hline
MWA5k-Hemisphere    &  $z=3.5$  & 0.061 & 0.11 & 0.54 & 0.39 & 0.84 & 0.4 & 0.11 & 0.87 & 0.41 \\ 
MWA5k-3 Fields      &  $z=3.5$  & 0.14  & 0.18 & 0.71 & 0.6 & 0.88 & 0.57 & 0.2 & 0.89 & 0.59 \\ 
FFTT                & $z=3.5$   & 0.02 & 0.054 & 0.32 & 0.29 & 0.8 & 0.19 & 0.067 & 0.87 & 0.31 \\  \hline
MWA5k-Hemisphere    &    $z=1.5$ & 0.055 & 0.1 & 0.58 & 0.45 & 0.88 & 0.47 & 0.11 & 0.89 & 0.46 \\            
MWA5k-3 Fields      &    $z=1.5$ & 0.12 & 0.17 & 0.75 & 0.68 & 0.88 & 0.69 & 0.19 & 0.89 & 0.69 \\

\end{tabular}        
\end{ruledtabular}                                                                                         
}        
\end{table*}

\begin{table*}
\tiny{
\caption{\label{minw} Constant Bias: 1$\sigma$ Errors on 
cosmological parameters with surveys centered at $z=1.5$ with $w$ as a free parameter.  
OPT refers to an
antenna temperature of 30K, PESS 100K, and 
CVL to the cosmic variance limited case.  
OPT and PESS correspond to what may
be possible by the time these experiments will 
be built and what is achievable today.}
\begin{ruledtabular}
\footnotetext[1]{When not otherwise indicated the ``OPT'' value of $T_{\rm sys}$ has been used.}
\begin{tabular}{lllllllllllll}
  & &     $\Omega_\Lambda$ & $\Omega_{\rm m}h^2$ & $\Omega_{\rm b}h^2$  & $n_{\rm s}$ & $A_{\rm s}^2 \times 10^{10}$ & $\alpha$ & $\Omega_\nu h^2$  & $w$ & $\tau$ & $Y_{\rm He}$ \\ \hline
Fiducial Values     &           & 0.7      & 0.147    & 0.023    & 0.95   & 25.0     & 0.0     & 0.00054 &-1.0 &0.10 & 0.24 \\ \hline
MWA5k-Hemisphere    &  PESS     &  0.0052 & 0.017 & 0.004 & 0.051 & -- & 0.019 & 0.003 & 0.047 & -- & -- \\
                    &  OPT      &  0.0047 & 0.012 & 0.0027 & 0.036 & -- & 0.014 & 0.0022 & 0.039 & -- & -- \\
                    &  CVL      &  0.0046 & 0.0088 & 0.0021 & 0.029 & -- & 0.012 & 0.0018 & 0.036 & -- & -- \\  \hline
MWA5k-3 Fields      &  PESS     &  0.012 & 0.039 & 0.0094 & 0.12 & -- & 0.045 & 0.0069 & 0.11 & -- & -- \\
                    &  OPT      &  0.011 & 0.027 & 0.0064 & 0.084 & -- & 0.033 & 0.0051 & 0.092 & -- & -- \\   
                    &  CVL      &  0.011 & 0.02 & 0.0048 & 0.068 & -- & 0.027 & 0.0043 & 0.083 & -- & -- \\   \hline
Planck              &           &  0.096 & 0.0061 & 0.00024 & 0.0094 & 0.27 & 0.0071 & 0.0059 & 0.16 & 0.0051 & 0.015 \\   
$+$MWA5k-Hemisphere &           &  0.0038 & 0.00056 & 0.00014 & 0.0042 & 0.22 & 0.0033 & 0.00049 & 0.023 & 0.0041 & 0.0065 \\
$+$MWA5k-3 Fields   &           &  0.0082 & 0.00097 & 0.00018 & 0.0064 & 0.22 & 0.0048 & 0.0009 & 0.045 & 0.0041 & 0.0097 \\
\end{tabular}                                                                                                                   
\end{ruledtabular}   
}                    
\end{table*}

\begin{table*}
\tiny{
\caption{\label{atan35} Fit to Scale Dependent Bias: 1$\sigma$ 
Errors on cosmological parameters with surveys 
centered at $z=3.5$.  OPT refers to an
antenna temperature of 30K, PESS 100K, 
and CVL to the cosmic variance limited case.  
OPT and PESS correspond to what may
be possible by the time these experiments 
will be built and what is achievable today.}
\begin{ruledtabular}
\footnotetext[1]{When not otherwise indicated the ``OPT'' value of $T_{\rm sys}$ has been used.}
\begin{tabular}{lllllllllllll}
  & &     $\Omega_\Lambda$ & $\Omega_{\rm m}h^2$ & $\Omega_{\rm b}h^2$  & $n_{\rm s}$ & $A_{\rm s}^2 \times 10^{10}$ & $\alpha$ & $\Omega_\nu h^2$   & $\tau$ & $Y_{\rm He}$ \\ \hline
Fiducial Values     &           & 0.7      & 0.147    & 0.023    & 0.95   & 25.0     & 0.0     & 0.00054    & 0.10 & 0.24 \\ \hline
MWA5k-Hemisphere    &  PESS    &   0.0043 & 0.011 & 0.0027 & 0.047 & -- & 0.028 & 0.002 & -- & --  \\
                    &  OPT     &   0.0025 & 0.0046 & 0.0011 & 0.028 & -- & 0.017 & 0.001 & -- & --  \\
                    &  CVL     &   0.00061 & 0.00026 & 0.000077 & 0.007 & -- & 0.0031 & 0.00033 & -- & --  \\  \hline
MWA5k-3 Fields      &  PESS     &  0.01 & 0.026 & 0.0062 & 0.11 & -- & 0.066 & 0.0047 & -- & --  \\
                    &  OPT      &  0.0058 & 0.011 & 0.0026 & 0.065 & -- & 0.04 & 0.0024 & -- & --  \\   
                    &  CVL      &  0.0014 & 0.0006 & 0.00018 & 0.016 & -- & 0.0072 & 0.00078 & -- & --  \\    \hline
FFTT                &           &  0.00079 & 0.00033 & 0.000089 & 0.0085 & -- & 0.0041 & 0.00035 & -- & --  \\     \hline
Planck              &           &  0.038    & 0.0041   & 0.00024  & 0.0094  & 0.25   & 0.007  & 0.0039  & 0.0046   & 0.014  \\
$+$MWA5k-Hemisphere &           &  0.0024 & 0.00052 & 0.00013 & 0.0042 & 0.22 & 0.0032 & 0.00051 & 0.0041 & 0.007 \\
$+$MWA5k-3 Fields   &           &  0.0055 & 0.00078 & 0.00017 & 0.0063 & 0.22 & 0.0048 & 0.00094 & 0.0041 & 0.01 \\ 
$+$FFTT             &           &  0.00079 & 0.00025 & 0.000078 & 0.0034 & 0.21 & 0.0017 & 0.00029 & 0.004 & 0.0052 \\ 
\end{tabular}                                                                                                                   
\end{ruledtabular}                                                                                                             
}                            
\end{table*}

\begin{table*}
\tiny{
\caption{\label{atan15} Fit to Scale Dependent Bias: 1$\sigma$ Errors on 
cosmological parameters with surveys centered 
at $z=1.5$.  OPT refers to an
antenna temperature of 30K, PESS 100K, 
and CVL to the cosmic variance limited 
case.  OPT and PESS correspond to what may
be possible by the time these experiments 
will be built and what is achievable today.}
\begin{ruledtabular}
\footnotetext[1]{When not otherwise indicated the ``OPT'' value of $T_{\rm sys}$ has been used.}
\begin{tabular}{lllllllllllll}
  & &     $\Omega_\Lambda$ & $\Omega_{\rm m}h^2$ & $\Omega_{\rm b}h^2$  & $n_{\rm s}$ & $A_{\rm s}^2 \times 10^{10}$ & $\alpha$ & $\Omega_\nu h^2$   & $\tau$ & $Y_{\rm He}$ \\ \hline
Fiducial Values     &           & 0.7      & 0.147    & 0.023    & 0.95   & 25.0     & 0.0     & 0.00054    & 0.10 & 0.24 \\ \hline
MWA5k-Hemisphere    &  PESS    &  0.0034 & 0.019 & 0.0047 & 0.067 & -- & 0.034 & 0.0032 & -- & --  \\
                    &  OPT     &  0.0024 & 0.013 & 0.0031 & 0.048 & -- & 0.025 & 0.0023 & -- & --  \\
                    &  CVL     &  0.0019 & 0.0095 & 0.0023 & 0.039 & -- & 0.02 & 0.0019 & -- & --  \\   \hline
MWA5k-3 Fields      &  PESS    &  0.008 & 0.045 & 0.011 & 0.16 & -- & 0.08 & 0.0076 & -- & --  \\
                    &  OPT     &  0.0056 & 0.03 & 0.0072 & 0.11 & -- & 0.058 & 0.0054 & -- & --  \\    
                    &  CVL      & 0.0043 & 0.022 & 0.0053 & 0.09 & -- & 0.046 & 0.0044 & -- & --  \\       \hline
Planck              &           &  0.038 & 0.0041 & 0.00024  & 0.0094  & 0.25   & 0.007  & 0.0039  & 0.0046   & 0.014  \\
$+$MWA5k-Hemisphere &           &  0.0021 & 0.00049 & 0.00014 & 0.0045 & 0.22 & 0.0037 & 0.00053 & 0.0041 & 0.0073 \\
$+$MWA5k-3 Fields   &           &  0.0048 & 0.00071 & 0.00018 & 0.0069 & 0.22 & 0.0052 & 0.00092 & 0.0042 & 0.011 \\ 
\end{tabular}                                                                                                                   
\end{ruledtabular}                                                                                                             
}                            
\end{table*}

\begin{table*}
\tiny{
\caption{\label{atanpoly} Polynomial Corrections to Bias: 1$\sigma$ Errors on 
cosmological parameters for FFTT centered around $z=3.5$.  Effects of fitting 
small polynomial corrections to the scale dependent bias.  Here n is the degree 
of the polynomial marginalized over.  The case with n=0 corresponds to
the value from table \ref{atan35}.  
}
\begin{ruledtabular}
\begin{tabular}{llllllll}
   &     $\Omega_\Lambda$ & $\Omega_{\rm m}h^2$ & $\Omega_{\rm b}h^2$  & $n_{\rm s}$ & $A_{\rm s}^2 \times 10^{10}$ & $\alpha$ & $\Omega_\nu h^2$  \\ \hline
Fiducial Values              & 0.7      & 0.147  & 0.023& 0.95& 25.0 & 0.0& 0.00054 \\ \hline
constant bias &   0.00076 & 0.00032 & 0.000087 & 0.0034 & -- & 0.0016 & 0.00032 \\ 
n=0           &   0.00079 & 0.00033 & 0.000089 & 0.0085 & -- & 0.0041 & 0.00035  \\
n=1           &   0.00084 & 0.00033 & 0.00009  & 0.011  & -- & 0.0071 & 0.00036  \\
n=2           &   0.00092 & 0.00035 & 0.000093 & 0.016  & -- & 0.014  & 0.00036  \\
n=3           &   0.00093 & 0.00037 & 0.000097 & 0.017  & -- & 0.015  & 0.00037  \\
n=4           &   0.00093 & 0.00038 & 0.0001   & 0.017  & -- & 0.015  & 0.00037  \\

\end{tabular}                                                                                                                   
\end{ruledtabular}                                                                                                             
}                            
\end{table*}

\section{Discussion}
The above results are very promising.  We find that using the
angular dependence of the power spectrum to isolate the term which 
only depends on cosmology cannot effectively constrain cosmological
parameters.  For this reason, it is crucial that the modulation of the
21cm power spectrum from the ionizing background is modeled reliably.

Our results show approximately how 
constraints would be degraded if a fit were to be necessary.
Table \ref{atanpoly} illustrates how more freedom in the fit affects 
cosmological constraints.  We find that the only parameters which 
are strongly degenerate with the new fitting parameters are $n_s$ and
$\alpha$, the spectral index and the running of the primordial
power spectrum.  The $1\sigma$ errors on these parameters could
increase by an order of magnitude.
The other $1\sigma$ cosmological parameter errors
are increased by roughly $20\%$ or less.
 
For comparison, the constraints from 21cm surveys that focus on the EoR
\cite{2008PhRvD..78b3529M} vary considerably under
different ionization power spectrum model assumptions.  
The robustness of our post-reionization
constraints is mainly due to the fact that we do not have to consider
the ionization power spectrum, only small scale dependent changes
to the bias.

Our results also depend
strongly on the maximum wavenumber of the power spectrum 
which can be used to reliably measure cosmological parameters, $k_{\rm max}$.
It will be desirable for theoretical predictions of the matter
power spectrum, presumably from N-body simulations, to be computed
to a fraction of a percent for $k_{max}$ values corresponding to
a $5 \%$ descrepancy in the linear and non-linear power spectra 
shown in Fig.\ \ref{cut}.  If the matter power spectrum can not be 
computed to this accuracy, the value of $k_{\rm max}$
may be the limiting factor for measuring 
cosmological parameters with experiments similar to those described in
this paper.  

The noise in the interferometers is lower at lower redshift due to
the redshift scaling of the galactic synchrotron emission. This
advantage is offset by the fact that at higher redshifts the neutral
fraction is higher and the observed power spectrum stays linear up to
a higher $k_{\rm max}$ value.  The FFTT and the MWA5000 at $z=1.5$
 are nearly cosmic variance limited.  These experiments would be close to capturing all
the information available from the power spectrum available to us on linear
scales at the relevant redshifts.  The MWA5000 with optimistic antenna temperature 
centered at $z=3.5$ gives errors roughly five times larger than the cosmic variance 
limited case.

Current cosmological information sets the upper limit on neutrino mass
to $\sim 0.62$ eV \cite{Goobar:2006xz,Seljak:2006bg}.  For FFTT with a 
constant bias, we
find a $1\sigma$ error corresponding to 0.030eV.  As
expected \cite{2008PhRvL.100p1301L}, this is clearly much better than
existing constraints and is sufficiently small to probe the neutrino
mass differences implied by neutrino oscillations.

It is also interesting to compare how well these surveys constrain
cosmology with respect to other future probes.  The FFTT survey will
 perform much better than
the Planck satellite for almost all of the parameters under
consideration.  When combined with Planck, essentially all of the
constraints are improved. In Table \ref{plancktable} we explicitly
show how the constraint on each parameter from Planck would improve if
combined with post-reionization 21cm surveys.  The most improved
parameters are $\Omega_\Lambda$, $\Omega_{\rm m}h^2$, and $\Omega_\nu h^2$,
which for FFTT are $2.0\%$, $5.4\%$, and 
$6.7\%$ of the errors from Planck alone.

Since most of the remaining neutral hydrogen after the EoR is found in
DLAs hosted by galaxies, we expect our results to be similar to 
galaxy redshift surveys over similar redshift ranges.  
In \cite{2008arXiv0805.1920P}
an all sky galaxy survey centered around $z=1$, termed G3, 
is considered.  We find the cosmological constraints 
they calculate to be very similar
to those in our MWA5000 case with a constant bias 
centered around $z=1.5$.
They also consider a Lyman break survey centered around $z=3$, which 
they term G2.  This survey contains roughly $1 \%$ of the volume 
included in our MWA5000-hemisphere and FFTT cases.  
Since the error on the power spectrum scales as the square root 
of the volume of a survey, this should cause
parameter errors in G2 to be roughly an order of magnitude larger.  We find
that this falls roughly into the range we calculate above for the constant
bias MWA5000 case centered around $z=3.5$.
Note that if the degeneracies due to the UV background discussed above
are present, galaxy surveys may do significantly better than 21cm 
at measuring the $n_s$ and $\alpha$, the spectral index of the primordial
power spectrum and its running.

In the case where we allow $w$ to vary as a free parameter we find the
constraints on most parameters essentially unchanged.  The main
exception is $\Omega_\Lambda$, where we find that our constraints are
increased by roughly a factor of two.  The constraints achievable for
$w$ are interesting, about $3.9\%$ if the whole hemisphere is observed,
and $9.2\%$ with only three fields of view.

\section{Conclusions}

A 21cm survey after reionization provides a promising probe of
cosmology and fundamental physics \cite{2008PhRvL.100p1301L}.  We find
the simple form of the 21cm power spectrum after reionization 
provides less uncertainty in forecasting constraints than 21cm during 
the EoR.  However, we find that errors on estimates of the spectral index 
and the running of the primordial
power spectrum could be increased by an order of magnitude
if the ionizing background introduces scale dependent effects into 
the post reionization 21cm power spectrum.  

Aside from its greater simplicity, the measurement sensitivity after
reionization is improved over EoR surveys due to lower foreground
brightness temperature, a bias that is greater than unity, and a larger
growth factor of density perturbations. These advantages are somewhat
offset by the fact that there is less neutral hydrogen after
reionization.

The constraints we derive are competitive with those from the EoR and
for the FFTT are
significantly better than next generation CMB experiments on their
own.  We find errors on $\Omega_\Lambda$, $\Omega_{\rm m}h^2$, and
$\Omega_\nu h^2$ to be improved the most, each lowering the errors from
Planck alone by at least an order of magnitude for both FFTT and the
survey similar in design to MWA5000.



\appendix*
\section{Modulation of 21cm PS from Ionizing Background}
We derive the form of the modulation of the 21cm power spectrum
due to the ionizing background in the case of a constant mean-free-path.  
The only difference in the 
derivation presented in \cite{2008arXiv0808.2323W} is the 
form of the fluctuations 
of the ionizing flux $\delta_J({\bf x})$.
We convolve the real-space density field
with a filter function to account for the effects of a finite
mean-free-path and the inverse squared dependence of ionizing flux.
The fluctuation in flux is given by

\begin{equation}
\delta_J({\bf x}) \propto \int d{\bf x'} G({\bf x},{\bf x'})\delta({\bf x'}).
\end{equation}
 
We assume a mean-free-path of ionizing photons which 
is independent of position such that,

\begin{equation}
G({\bf x},{\bf x'}) \propto 
\frac{\exp(-|{\bf x'}-{\bf x}|/\lambda_{\rm mfp})}{|{\bf x'}-{\bf x}|^2},
\end{equation}

where $\lambda_{\rm mfp}$ is the mean free path of ionizing photons.
The convolution theorem gives

\begin{equation}
\delta_J({\bf k}) \propto g({\bf k})\delta_k,
\end{equation}
where $g({\bf k})$ and $\delta_k$ are the Fourier transforms
of the filter function and the fluctuation in ionizing flux.
We find that 
\begin{equation}
g({\bf k})  \propto \arctan(k\lambda_{\rm mfp})/k.
\end{equation}

From the derivation presented in \cite{2008arXiv0808.2323W},
this implies that the power spectrum is given by Eqn. \ref{PSeqn}
with  

\begin{equation}
K(k) = K_0\arctan(k\lambda_{\rm mfp})/\lambda_{\rm mfp}k,
\end{equation}
where $K_0$ is a constant of the 
order of $0.01$ or smaller \cite{2008arXiv0808.2323W}.

\begin{acknowledgments}
This work was supported in part by a NASA LA grant and by Harvard
University funds.  We thank Matt McQuinn and Jonathan Pritchard for
useful discussions.
\end{acknowledgments}

\bibliography{paper}

\end{document}